\pdfoutput=1

\documentclass[11pt,twoside,a4paper,cmspaper,final,collab]{cms-tdr}
\def\svnVersion{69e6a1e-D}\def\svnDate{2020/07/21}\def\cmsCernNoTag{CERN-EP-2020-032}\def\cmsCernDate{\today}\def\cmsMessage{"Published in the Journal of High Energy Physics as \href{http://dx.doi.org/10.1007/JHEP07(2020)116}{\doi{10.1007/JHEP07(2020)116}.}"}

\begin{document}\cmsNoteHeader{HIN-18-016}

\hyphenation{had-ron-i-za-tion}
\hyphenation{cal-or-i-me-ter}
\hyphenation{de-vices}
\newcommand {\jetphox} {{\textsc{jetphox}}\xspace}
\newcommand {\ngr} {\ensuremath{N^{\PGg}_\text{raw}}\xspace}
\newcommand {\ngc} {\ensuremath{N^{\PGg}_\text{corrected}}\xspace}
\newcommand {\ngu} {\ensuremath{N^{\PGg}_\text{unfolded}}\xspace}
\newcommand {\pp} {\ensuremath{\Pp\Pp}\xspace}
\newcommand {\pPb} {\ensuremath{\Pp\mathrm{Pb}}\xspace}
\newcommand {\ptg} {\ensuremath{\pt^\PGg}\xspace}
\newcommand {\etg} {\ensuremath{\ET^\PGg}\xspace}
\newcommand {\TAA} {\ensuremath{T_\mathrm{AA}}\xspace}
\newcommand {\nmb} {\ensuremath{N_\mathrm{MB}}\xspace}
\newcommand {\raa} {\ensuremath{R_{\mathrm{AA}}}\xspace}
\newcommand {\ptGen} {\ensuremath{\pt^\text{gen}}\xspace}
\newcommand {\sumIso} {\ensuremath{I}\xspace}
\newcommand {\sumIsoGen} {\ensuremath{I^\text{gen}}\xspace}
\newcommand {\ztoee} {\ensuremath{\PZ\to\Pep\Pem}\xspace}
\newcommand {\pythiahydjet} {\PYTHIA{}{+}\HYDJET{}\xspace}

\providecommand{\NA} {\ensuremath{\text{---}}}
\newlength\cmsTabSkip\setlength{\cmsTabSkip}{1ex}

\cmsNoteHeader{HIN-18-016}
\title{The production of isolated photons in PbPb and \texorpdfstring{\pp}{pp} collisions at $\sqrtsNN = 5.02\TeV$}

\date{\today}

\abstract{The transverse energy (\etg) spectra of photons isolated from other particles are measured using proton-proton (\pp) and lead-lead (PbPb) collisions at the LHC at $\sqrtsNN = 5.02\TeV$ with integrated luminosities of 27.4\pbinv and 404\mubinv for \pp and PbPb data, respectively. The results are presented for photons with $25 < \etg < 200\GeV$ in the pseudorapidity range $\abs{\eta} < 1.44$, and for different centrality intervals for PbPb collisions. Photon production in PbPb collisions is consistent with that in \pp collisions scaled by the number of binary nucleon-nucleon collisions, demonstrating that photons do not interact with the quark-gluon plasma. Therefore, isolated photons can provide information about the initial energy of the associated parton in photon+jet measurements. The results are compared with predictions from the next-to-leading-order \jetphox generator for different parton distribution functions (PDFs) and nuclear PDFs (nPDFs). The comparisons can help to constrain the nPDFs global fits.
}

\hypersetup{
pdfauthor={CMS Collaboration},
pdftitle={The production of isolated photons in PbPb and pp collisions at sqrt(s[NN]) = 5.02 TeV},
pdfsubject={CMS},
pdfkeywords={CMS, physics, heavy-ions, photon}}

\maketitle

\section{Introduction}
One of the most important reasons for studying relativistic heavy 
ion collisions is understanding 
the deconfined state of matter, so called quark-gluon plasma (QGP), which is predicted 
by the theory of strong 
interactions, quantum chromodynamics (QCD), to exist at high 
temperatures and energy density~\cite{Karsch:2003jg,Bazavov:2011nk,Borsanyi:2010bp,Busza:2018rrf}. 
In heavy ion collisions, the expectation is that 
high transverse momentum (\pt) photons do not 
strongly interact with the QGP and thus provide a direct way to 
test perturbative QCD (pQCD). 
Comparing photon production in proton-proton (\pp) and heavy ion collisions 
is important to both establish that we understand the production of 
photons in collisions of nuclei and that the photons are 
not affected by the medium through which they pass. 
In contrast to photons, partons lose energy in the medium and 
their production is significantly modified compared to \pp collisions~\cite{Khachatryan:2016odn,Sirunyan:2018nsz,Sirunyan:2017isk}.
The production of photons paired back-to-back with jets from fragmented partons has been studied 
at the CERN LHC~\cite{Sirunyan:2018ncy,Sirunyan:2018qec,Sirunyan:2017qhf,Aaboud:2018anc} 
to test energy loss in the strongly interacting medium produced in heavy ion 
collisions.

Prompt photons are defined to be those produced directly 
from the hard scattering of two partons, or 
fragmented collinearly from final-state partons at high-\pt~\cite{Ichou:2010wc}.
At leading order (LO), partons produce photons through two hard scattering subprocesses: 
Compton scattering $\PQq\Pg \to \PQq\PGg$ and 
quark-antiquark annihilation $\qqbar \to \Pg\PGg$, of 
which Compton scattering is dominant~\cite{Ichou:2010wc}.
To identify photons from parton scattering requires that the photons be isolated from 
other particles in order to reduce  
a large background of decay photons 
coming from neutral mesons (mostly $\PGpz\to\PGg\PGg$).
This isolation requirement also suppresses the contribution from 
fragmentation processes~\cite{Ichou:2010wc}. 
As a result, isolated photon production is sensitive to the 
gluon parton distribution functions (PDFs).

The scaled ratio of the production cross sections 
in \pp and heavy ion collisions is known as the nuclear modification factor, 
\begin{equation}
\raa(\pt) = \frac{1}{\TAA}\frac{1}{\nmb}\frac{\rd N^\mathrm{AA}/\rd \pt}{\rd \sigma^{\pp}/\rd \pt},
\end{equation}
where \nmb is the number of sampled minimum-bias (MB) events 
in nucleus-nucleus (AA) collisions, 
and \TAA is the nuclear overlap function~\cite{Miller:2007ri}, which is given by 
the number of binary nucleon-nucleon (NN) collisions divided by 
the inelastic NN cross section.
This \TAA can be interpreted as the NN-equivalent integrated 
luminosity per heavy ion collision. 
Here, $\rd N^\mathrm{AA}/\rd \pt$ is the yield in AA collisions in a \pt interval 
and $\rd \sigma^{\pp}/\rd \pt$ is the differential cross section
in inelastic \pp collisions.
A value of $\raa=1$ indicates that PbPb collision data are compatible with a superposition of \pp 
collisions, while a deviation from unity indicates
either enhancement or suppression of isolated photon production.
The \raa of isolated photons allows 
an estimation of possible modification of the PDFs in a nucleus 
compared to a simple incoherent superposition of nucleon PDFs~\cite{deFlorian:2003qf,Hirai:2007sx}.
A typical form of such modifications is to have suppression 
at low Bjorken $x \lesssim 10^{-2}$ (shadowing), and enhancement at $x \sim 10^{-1}$ (anti-shadowing)~\cite{Arleo:2011gc}.

The differential cross section for isolated photons was extensively studied
at the LHC in \pp collisions at various collision
energies~\cite{Sirunyan:2018gro,Chatrchyan:2011ue,Khachatryan:2010fm,Aad:2016xcr,Aad:2013zba,Acharya:2019jkx}.
In heavy ion collisions, measurements of \raa for isolated photons were performed
in lead-lead (PbPb) collisions at a center-of-mass 
energy per nucleon pair $\sqrtsNN = 2.76\TeV$ with 
the CMS~\cite{Chatrchyan:2012vq} and ATLAS~\cite{Aad:2015lcb} detectors, 
and in proton-lead (\pPb) collisions at $\sqrtsNN = 8.16\TeV$ with 
the ATLAS detector~\cite{Aaboud:2019tab}.
The ALICE Collaboration reported similar measurements in PbPb collisions at
$\sqrtsNN = 2.76\TeV$~\cite{Adam:2015lda} at a lower \pt range than that 
used in the CMS and ATLAS measurements. 
In the \pPb and PbPb LHC measurements, it was found that the 
production of high-\pt prompt photons is not significantly modified
by the medium and is compatible with the pQCD calculations.

In this paper, measurements of the differential cross 
sections for isolated photons
in \pp and PbPb collisions, as well as the nuclear modification factors 
of isolated photons, are reported at $\sqrtsNN = 5.02\TeV$, using 
data taken in 2015 with the CMS detector. 
The measurements are performed over the photon transverse 
energy ($\etg \equiv \ptg c$) range of 
$25 < \etg < 200 \GeV$ 
for the photon pseudorapidity $\abs{\eta} < 1.44$.
This \etg range 
corresponds to the 
kinematic region of $0.01 < x_\mathrm{T} < 0.08$, where $x_\mathrm{T} = 2\etg/\sqrtsNN$.
Both shadowing and anti-shadowing effects are expected in this region. 
The measurements are compared with the pQCD next-to-leading order (NLO) calculations from 
\jetphox~\cite{Aurenche:2006vj} with free proton PDFs and nuclear PDFs (nPDFs). 
The present results can be used in a global fit analysis of nPDFs 
to constrain gluon parton densities in nuclei.
In addition, the 
current measurements provide baselines to find any modification of initial 
parton states by the nuclear medium for jet events tagged by isolated photons.
These data, which represent the first measurement of isolated photons 
for PbPb collisions at $\sqrtsNN = 5.02\TeV$, have a much higher 
statistical significance and a larger \etg range than the 
previous measurement in PbPb collisions 
at $\sqrtsNN = 2.76\TeV$~\cite{Chatrchyan:2012vq,Aad:2015lcb}.

\section{The CMS detector} \label{sec:CMS}
The central feature of the CMS detector system is a superconducting solenoid of
6\unit{m} internal diameter, providing a magnetic field of 3.8\unit{T}. Within
the solenoid volume are silicon pixel and strip trackers, which measure
the charged-particle trajectories within the range of $\abs{\eta} < 2.5$, a lead
tungstate crystal electromagnetic calorimeter (ECAL), and a brass and
scintillator hadron calorimeter (HCAL). Each detector element consists of a barrel and two
endcap sections. The barrel and endcap calorimeters provide $\abs{\eta}$ coverage
out to 3. 

The photon candidates used in this analysis are reconstructed using the
energy deposited in the barrel region of the ECAL, which covers
$\abs{\eta} < 1.442$. 
In the barrel section of the ECAL, an energy resolution of 
about 1\% is achieved for unconverted or late-converting photons that 
have energies in the range of tens of GeV. The remaining barrel photons 
have a resolution of about 1.3\% up to $\abs{\eta} = 1$, rising 
to about 2.5\% at $\abs{\eta} = 1.4$~\cite{Khachatryan:2015iwa}. 

The hadron forward (HF) calorimeters extend the $\abs{\eta}$
coverage of the HCAL to $\abs{\eta} = 5.2$. 
Each HF calorimeter consists of 432 readout towers, containing long and short 
quartz fibers running parallel to the beam. The long fibers run the entire depth 
of the HF calorimeter (165\unit{cm}, or approximately 10 interaction lengths), while 
the short fibers start at a depth of 22\unit{cm} from the front of the detector. 
By reading out the two sets of fibers separately, it is possible to distinguish 
showers generated by electrons and photons, which deposit a large fraction of 
their energy in the long-fiber calorimeter segment, from those generated by 
hadrons, which produce on average nearly equal signals in both calorimeter segments.
In PbPb collisions, the HF calorimeters are used
to determine the centrality of the collision,
which is defined by the geometrical overlap of the two 
colliding Pb nuclei~\cite{Chatrchyan:2011sx}. Muons are detected in
gas-ionization chambers embedded in the steel flux-return yoke outside the
solenoid. 

Events of interest are selected using a two-tiered trigger system~\cite{Khachatryan:2016bia}. 
The first level (L1), composed of custom hardware processors, uses information from 
the calorimeters and muon detectors to select events at a rate of around 100\unit{kHz} 
within a time interval of less than 4\mus. The second level, known as the high-level 
trigger (HLT), consists of a farm of processors running a version of the full event 
reconstruction software optimized for fast processing, and reduces the event rate 
to around 1\unit{kHz} before data storage.

A more detailed description of the CMS detector, together with
a definition of the coordinate system used and the relevant kinematic
variables, can be found in Ref.~\cite{Chatrchyan:2008zzk}.

\section{Analysis procedure} \label{sec:analysis}
\subsection{Monte Carlo simulation}
\label{sec:MC}
Simulated Monte Carlo (MC) events samples of \pp collisions
are generated with \PYTHIA 8.212 \cite{Sjostrand:2014zea} 
using tune CUETP8M1~\cite{Khachatryan:2015pea}. 
For PbPb collisions,
\PYTHIA events are embedded into events generated with \HYDJET 1.8~\cite{Lokhtin:2005px},
which is tuned to reproduce global event properties such as the charged-hadron
\pt spectrum and particle multiplicity. 
The prompt photon, dijet, and \ztoee events are used 
in corrections for detector effects and background rejection. 
The generated events are propagated through the full CMS detector
using the \GEANTfour simulation package~\cite{geant4}.
The energy of photon candidates in simulations is smeared to account for 
the difference in photon energy resolution between data and simulations.

\subsection{Event selection}
\label{sec:eventSelection}
Events with photons are selected from photon-dedicated triggers.
Offline, several event selection criteria are used to remove non-hadronic events in \pp and PbPb collisions.
Events are required to contain at least one reconstructed vertex with at least
two tracks within the vertex $z$ position range of $\abs{z}< 15\unit{cm}$. This requirement removes
noncollision background events such as beam-gas interactions or beam scraping
events near the interaction point~\cite{Khachatryan:2016odn,Sirunyan:2017qhf}.
Additionally, at least three detector elements with energies greater 
than 3\GeV in the HF on each side of 
the interaction point are required in PbPb events.
This condition rejects most of the electromagnetic interactions from
ultra-peripheral heavy ion collisions.
In PbPb collisions, the cluster shapes of the silicon pixel detector are required to be compatible with the vertex position.  

The event selection efficiency in PbPb collisions is $(99 \pm 2)\%$.
This number can be above 100\% because of remaining contamination 
from electromagnetic interactions
in the selected event sample~\cite{Djuvsland:2010qs}.
The efficiency-corrected $\nmb$ for the 0--100\% centrality range is $2.72\times 10^9$, 
corresponding to a total integrated luminosity of 404\mubinv.
The total integrated luminosity of the \pp event sample is 27.4\pbinv with
an uncertainty of $2.3\%$~\cite{CMS-PAS-LUM-16-001}.

In PbPb collisions, the event centrality is estimated by the measured fraction of the total
inelastic hadronic cross section. The percentage starts from 0\% for the most central collisions,
with the smallest impact parameter and the largest nuclear
overlap, and goes to 100\% for the most peripheral collisions.
Such peripheral collisions are the closest to a \pp-like environment~\cite{Chatrchyan:2011sx}.

Results of this analysis are presented in four centrality intervals:
0--10\% (most central), 10--30\%, 30--50\% and 50--100\% (most peripheral).
The \TAA values are determined from a Glauber model calculation~\cite{Miller:2007ri}, and 
their averages are listed in Table~\ref{tab:TAA} for the four centrality bins.
Uncertainties in \TAA are estimated by varying the Glauber model parameters~\cite{Khachatryan:2016odn}. 

\begin{table}[tbh]
\centering
\topcaption{Average numbers of the nuclear overlap function ($\langle\TAA\rangle$) and
their uncertainties for various centrality ranges used in this analysis.}
\newcolumntype{x}{D{,}{\text{--}}{2.3}}
\newcolumntype{z}{D{,}{}{5.5}}
\setlength\extrarowheight{1.5 pt}
\begin{tabular}{xzz}
 \multicolumn{1}{c}{Centrality} &  \multicolumn{1}{c}{$\langle\TAA\rangle$ [mb$^{-1}$]} \\
\hline
0,100\%  & 5.61,{^{+0.16}_{-0.19}} \\[1.5pt]
0,10\%  & 23.22,{^{+0.43}_{-0.69}} \\[1.5pt]
10,30\%  & 11.51,{^{+0.30}_{-0.39}} \\[1.5pt]
30,50\%  & 3.82,{^{+0.21}_{-0.21}} \\[1.5pt]
50,100\%  & 0.44,{^{+0.05}_{-0.03}} \\[1.5pt]
\end{tabular}
\label{tab:TAA}
\end{table}

\subsection{Photon reconstruction and identification}
\label{sec:photonRecoID}
Two different dedicated photon triggers are used
in this analysis. For photons with $\etg > 40\GeV$,
candidates are selected online by L1 triggers by requiring an ECAL transverse energy
deposit larger than 21 (20)\GeV in PbPb (\pp) collisions.
For photons with $20 < \etg < 40 \GeV$, all MB events are used
for L1 trigger selection in PbPb collisions, 
which requires a coincidence of signals above threshold in 
both sides of the HF calorimeters.
Events with an ECAL transverse energy
deposit larger than 5\GeV are selected by the L1 trigger in \pp collisions.
The preselected photons are reconstructed by the HLT using
the ``island'' clustering algorithm in PbPb collisions, and
the ``hybrid'' clustering algorithm in \pp collisions~\cite{Khachatryan:2015iwa,Chatrchyan:2012vq}.
Events with at least one reconstructed photon of 
$\etg > 40$ (20)\GeV
are selected by the HLT for high- (low-)\etg photons.
The HLT selections of both triggers are found to be fully efficient
for photons in PbPb events,
while the HLT triggers for photons in \pp events  
are inefficient up to 5\GeV above the thresholds of 40 (20)\GeV for 
high- (low-)\etg photons.
Photons in \pp collisions are reconstructed offline with the ``Global Event Description (GED)'' algorithm
detailed in Ref.~\cite{Khachatryan:2015iwa}, while the ``island'' clustering algorithm
is used in PbPb collisions, which is optimized for high-multiplicity
PbPb events as described in Ref.~\cite{Chatrchyan:2012vq}.

In order to reject electrons in $\abs{\eta} < 1.442$ that are misidentified as photons, the photon candidates
are discarded if the differences in $\eta$ or 
azimuthal angle ($\phi$, in radians) between the photon
candidate and any electron candidate track with $\pt > 10\GeVc$ are less than 0.03.~\cite{Chatrchyan:2012vq}.
Anomalous signals caused by highly ionizing particles interacting directly 
with the silicon avalanche photodiodes in the ECAL barrel readout are removed
using the prescription given in Ref.~\cite{Chatrchyan:2012vq}.

The energy of the reconstructed photons is corrected to account for 
the effects of the material in
front of the ECAL and for the incomplete containment of the 
shower energy~\cite{Khachatryan:2015iwa}.
To account for underlying event (UE) contamination from soft collisions
in PbPb data, corrections obtained from the simulation using \PYTHIA and
\pythiahydjet photon events are applied.

Only photon candidates with the ratio of HCAL over
ECAL energies ($H/E$) less than 0.1 inside a cone of radius $\Delta R = \sqrt{\smash[b]{(\Delta\eta)^2+(\Delta\phi)^2}} = 0.15$ around the photon
candidate are selected to reject high-\pt hadrons. The remaining background 
contributions from decay photons
are suppressed by imposing the isolation requirement, resulting in a sample
enriched in prompt photons.
The generator-level isolation (\sumIsoGen) is defined as
the $\ET^\text{gen}$ sum  
of all the other final-state particles, excluding 
neutrinos, in a cone of radius
$\Delta R = 0.4$ around the photon candidates.
The isolation variable (\sumIso) for a reconstructed photon is 
given by the sum of transverse energies in ECAL and HCAL and 
the transverse momenta of all tracks with $\pt > 2\GeVc$ in trackers 
inside the cone of $\Delta R = 0.4$ around the photon candidates. 
The UE is corrected when measuring \sumIso in PbPb data
by subtracting the average value of the energy in a rectangular area
with length of $2\Delta R$ in the $\eta$-direction around a photon candidate and width of $2\pi$ in the $\phi$-direction,
while no UE correction is applied in \pp data.
An \sumIso value less than 1\GeV is required for reconstructed photon candidates, which 
corresponds to an \sumIsoGen value less than 5\GeV for generated photons.
This tightened criterion of $\sumIso<1\GeV$ compared to $\sumIsoGen<5\GeV$ is optimized 
to minimize the impact of UE fluctuations from studying the correlations of
\sumIso and \sumIsoGen in \PYTHIA and \pythiahydjet samples.
More detailed descriptions can be found in Ref.~\cite{Chatrchyan:2012vq}.

After applying $H/E$ and isolation requirements, the dominant background
photons come from the contribution from isolated neutral mesons, \eg, \PGpz, \PGh, and
\PGo, decaying into two or three closely spaced photons and 
misidentified as a single isolated photon.
This background can be significantly reduced by a requirement 
on the shower shape, which is a measure of
how energy deposited in the ECAL is distributed in $\phi$ and $\eta$.
The electromagnetic shower shape variable $\sigma_{\eta\eta}$ is
defined as a modified second moment of the ECAL energy cluster
distribution around its mean $\eta$ position~\cite{Khachatryan:2010fm,AWES1992130}:

\ifthenelse{\boolean{cms@external}}{
    \begin{multline}
    \label{sieieFormula}
    \sigma_{\eta\eta}^2 = \frac{\sum_i^{5{\times}5}w_i(\eta_i-\eta_{5{\times}5})^2}{\sum_i^{5{\times}5} w_i}, \\
         w_i = \max\left(0, 4.7 + \ln \frac{E_i}{E_{5{\times}5}}\right),
        \end{multline}
}{
    \begin{equation}
    \label{sieieFormula}
    \sigma_{\eta\eta}^2 = \frac{\sum_i^{5{\times}5}w_i(\eta_i-\eta_{5{\times}5})^2}{\sum_i^{5{\times}5} w_i}, \qquad w_i = \max\left(0, 4.7 + \ln \frac{E_i}{E_{5{\times}5}}\right).
        \end{equation}
}
Here $E_i$ and $\eta_i$ are the energy deposit and $\eta$ of the $i$th ECAL
crystal within a $5{\times}5$ crystal array centered around the electromagnetic
cluster, and $E_{5{\times}5}$ and $\eta_{5{\times}5}$ are the total energy and
mean $\eta$ of the $5{\times}5$ crystal matrix, respectively.
Photon candidates are required to have $\sigma_{\eta\eta}$ less 
than 0.01 since most decay photons have larger values of $\sigma_{\eta\eta}$. 
Thus, this cut further enriches the fraction of prompt photons in the sample.

\subsection{Signal extraction}
\label{sec:pho_sig_corr}
\begin{figure*}[h]
  \centering
    \includegraphics[width=0.55\textwidth]{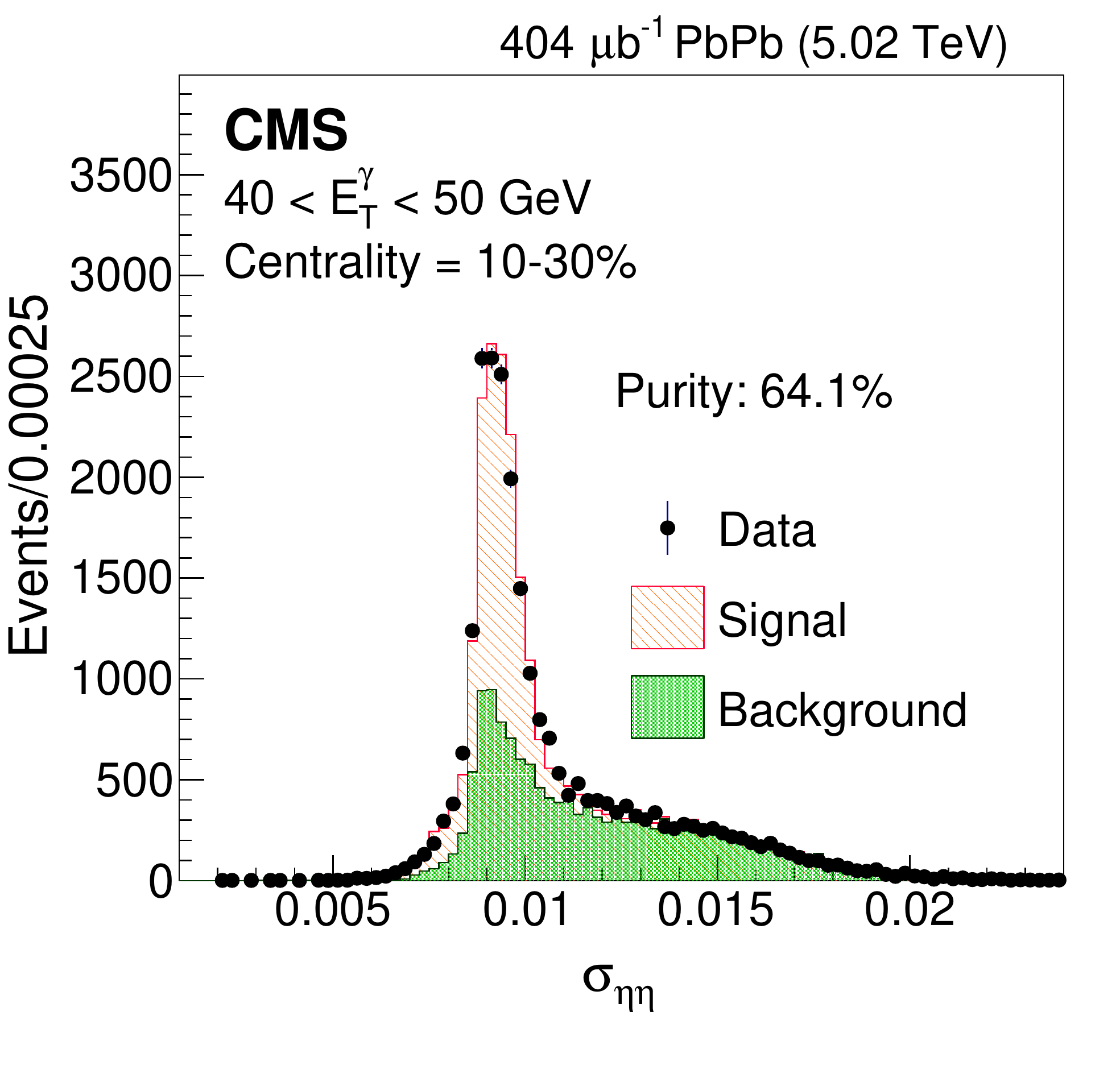}
    \caption{Template fit of the shower shape variable
      $\sigma_{\eta\eta}$ for $40 < \etg < 50\GeV$ in the 10--30\% centrality class. The black
      points show the PbPb experimental data. The red histogram is the
      signal template obtained from \pythiahydjet simulations, and the green histogram
      is the background template estimated from the data for the nonisolated sideband region.
	  Purity values are estimated in the range of $\sigma_{\eta\eta}< 0.01$.}
    \label{fig:PbPb-purity}
\end{figure*}

After the selection conditions are applied, the 
remaining backgrounds of decay photons from hadrons are
estimated by using a two-component template fit of $\sigma_{\eta\eta}$.
The signal template is obtained from simulations, and the background
shape is obtained from the data in a nonisolated sideband region ($1 < \sumIso < 5\GeV$).
The sideband region is chosen to be close to the signal region
in order to reduce bias from the correlation between $\sigma_{\eta\eta}$ and 
\sumIso. The signal contamination in the sideband region is estimated by taking 
the signal shape from simulation and normalizing with the fraction 
between the signal and the sideband regions. 
The normalized signal 
shape is then subtracted from the background template. 
The purity, which is the fraction of prompt photons within the remaining
candidates, is determined from the template fit. An example is shown
in Figure~\ref{fig:PbPb-purity} for the photons with $40 < \etg < 50\GeV$
in the 10--30\% centrality class. The purity decreases in more
central collisions, reflecting an increase in background contributions.
The raw signal yield (\ngr) is defined as the number 
of photon candidates passing all selection criteria. In 
order to correct for the remaining background, \ngr is reduced 
by the purity factor obtained from the template fits.

\subsection{Efficiency corrections}
\label{sec:pho_eff_corr}
\begin{figure*}[ht]
  \centering
    \includegraphics[width=0.95\textwidth]{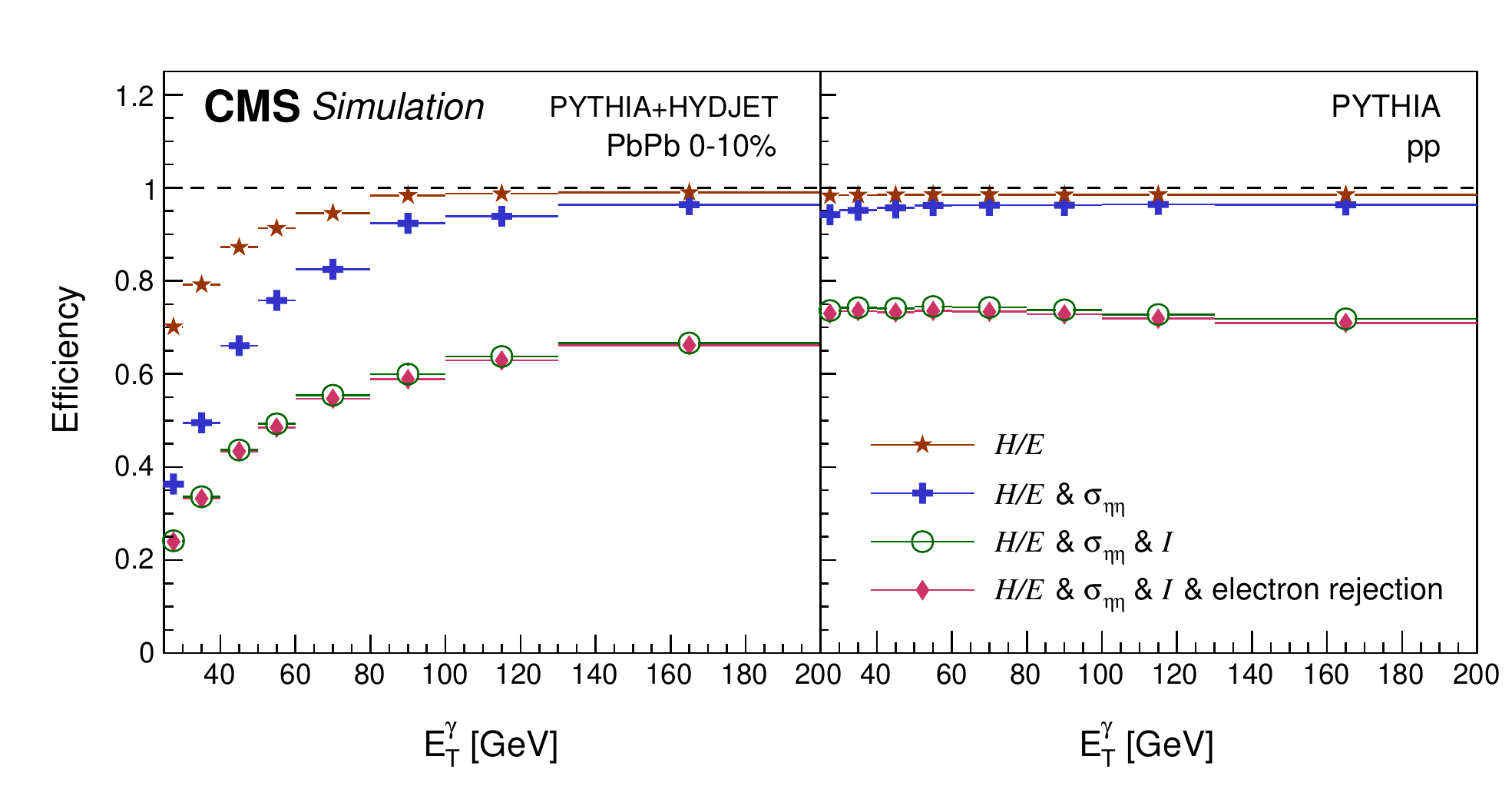}
    \caption{Efficiency of the isolated photon detection as a function of \etg 
        for PbPb collisions in the 0--10\% centrality range (left) and for \pp data (right).
            The different colors represent various selection criteria: $H/E<0.1$, $\sigma_{\eta\eta}< 0.01$, $\sumIso<1\GeV$ and electron rejection criterion.}
    \label{fig:efficiency}
\end{figure*}

The efficiency to detect isolated photons using different reconstruction 
selection criteria is extracted from
simulations as a function of \etg. Figure~\ref{fig:efficiency} shows the 
signal efficiency obtained from \pythiahydjet and \PYTHIA 
for 0--10\% centrality PbPb and for \pp collisions, respectively. 
The total efficiency is obtained by multiplying signal selection,
trigger, and reconstruction efficiencies. The reconstruction efficiency 
is calculated from simulations as the ratio of reconstructed photon candidates 
by the reconstruction algorithms (``island'' for PbPb and 
``GED'' for \pp collisions) to generated photons. 
The reconstruction efficiency is about 
99.0 and 99.5\% for \pp and PbPb collisions, respectively, 
for all \etg ranges, showing no centrality dependence.
The trigger efficiency is obtained from the data.
The scale factors (SF), the efficiency ratio of data to simulations, are estimated 
with \ztoee events using the
``tag-and-probe'' method~\cite{Khachatryan:2015iwa} by matching 
electrons to photon candidates.
The SF are applied to the total 
efficiency to account for the efficiency difference between the data and simulation. 
The total efficiency is applied as a correction to the \ngr values.

\subsection{Unfolding}
\label{sec:pho_unfolding}
The photon signal yields corrected by efficiency and purity can be described as 
\begin{equation}
\ngc = \frac{\ngr P}{\epsilon}, 
\end{equation}
where $\epsilon$ is the total efficiency, and $P$ is the purity correction factor.
The \ngc are unfolded for detector resolution. 
Response matrices are constructed from \pythiahydjet (\PYTHIA) for 
PbPb (\pp) data in different centrality bins.
A matrix inversion method is used without regularization 
in the \textsc{RooUnfold} software package~\cite{Adye:2011gm}.
The unfolded spectra (\ngu) are used in the cross section determination.

\subsection{Systematic uncertainties}
\label{sec:sys}
The systematic uncertainties are summarized in Table~\ref{table:SysXsec}
for the cross section of isolated photons in \pp and PbPb collisions, 
and in Table~\ref{table:SysRAA} for the nuclear modification factors of isolated photons.
All systematic uncertainties are evaluated 
by varying the quantity relevant to each source and 
propagating the change to the final observables, 
and then taking the deviation from the nominal results.
The total uncertainty is obtained as the quadratic sum of systematic uncertainties from the different sources.
The systematic uncertainties from most of the sources partially cancel in the \raa analysis
because the systematic variations are applied to both \pp and PbPb data.

\begin{table*}[htbp]
 \centering
 \topcaption{\label{table:SysXsec} Summary of the contributions from various sources to the estimated systematic
   uncertainties in the cross section of isolated photons in \pp and PbPb collisions.
 When ranges are shown, they indicate the \etg-dependent variations of the uncertainties.}
 \begin{tabular}{lcccccc}
 & \pp &   \multicolumn{5}{c}{PbPb centrality} \\
   Source & & 0--100\% & 0--10\% & 10--30\% & 30--50\%   & 50--100\%  \\
   \hline
    Purity & 4--15\% & 5--15\%  & 9--16\%  & 11--14\% & 5--18\%  & 5--17\%  \\
    Electron rejection & $<$0.4\% &  1--3\% &  1--10\%  & 1--5\% & 1--3\%  & 0--7\% \\
    Pileup   & 0--11\% \  & \NA & \NA & \NA & \NA & \NA \\ 
    Energy scale  & 1--2\% & 3--8\% & 2--7\% & 2--10\% & 2--11\%   & 1--12\%  \\
    Energy resolution & $<$0.2\% & 1--3\%  & 1--7\% & 1--9\% & 1--8\%  & 2--6\% \\
    Unfolding   & $<$0.2\% & 1--4\% &  0--9\%  & 0--5\% & 0--3\%  & 0--1\%\\
    Efficiency & 1--2\% & 0--1\% & 0--4\% & 0--2\% & 0--1\%  & 0--3\%  \\[\cmsTabSkip]
    Integrated luminosity & 2.3\% \ & \NA & \NA & \NA & \NA & \NA \\
    \TAA & \NA &  4\%  & 3\% &  4\% &  6\%  &  11\% \\[\cmsTabSkip]
    Total  & 4--16\%  & 6--18\%  & 14--21\% &  12--18\% &  10--20\%  &  10--21\% \\
 \end{tabular}
\end{table*}

\begin{table*}[htbp]
 \centering
 \topcaption{\label{table:SysRAA}  Summary of the contributions from various sources to the estimated systematic
   uncertainties in the nuclear modification factors calculated from \pp and PbPb data.
 When ranges are shown, they indicate the \etg-dependent variations of the uncertainties.}
 \begin{tabular}{lccccc}
 & \multicolumn{5}{c}{PbPb centrality} \\
   Source & 0--100\% & 0--10\% & 10--30\% & 30--50\%  & 50--100\%  \\
   \hline
    Purity & 6--9\%  & 7--13\%  & 3--12\% & 4--8\%  & 2--7\%  \\
    Electron rejection & 1--2\%  & 0--10\% & 1--6\% & 0--3\% & 0--7\% \\
    Pileup & 0--10\%  & 0--10\%  & 0--10\% & 0--10\%  & 0--10\% \\
    Energy scale  & 2--4\% & 3--6\% & 1--9\%  & 2--7\%  & 1--10\%  \\
    Energy resolution & 0--3\% & 1--7\% & 0--9\% & 1--8\%  & 2--6\% \\
    Unfolding & 1--4\%  & 1--9\%  & 1--5\% & 0--3\% & 0--1\% \\
    Efficiency & 0--2\% & 0--5\% & 0--2\% & 0--1\% & 0--2\% \\[\cmsTabSkip]
    Integrated luminosity & 2.3\% \ & 2.3\% \ & 2.3\% \ & 2.3\% \ & 2.3\% \\
    \TAA & 4\% & 3\% & 4\% & 6\% & 11\% \\[\cmsTabSkip]
    Total   & 5--12\%   & 10--17\% &  6--18\% &  7--15\%  &  7--15\% \\
 \end{tabular}
\end{table*}

One of the dominant sources of systematic uncertainty is the purity determination.
The sideband definition used for producing the background template is changed to
tight ($1 < \sumIso < 3\GeV$) or loose ($5 < \sumIso < 10\GeV$) nonisolated selection 
criteria to evaluate this uncertainty.

After the electron rejection process, there are still electrons which 
are misidentified as photons.
The rejection rate is calculated from simulations, and the remaining 
number of misidentified electrons is subtracted 
from the \ngr values as an additional correction 
for the systematic uncertainty of electron rejection.
The difference between the nominal and subtracted \ngr values
are propagated to the final results and quoted as systematic uncertainty. 

Pileup events have multiple interactions within a recorded event with corresponding 
multiple primary vertices. For PbPb collisions, the 
effect of pileup events on the photon spectra is negligible. 
The systematic uncertainty from the pileup contribution in \pp collisions is estimated 
by counting \ngr when the number of primary vertices in the events is one. 

The mean and width of the invariant mass distribution of \PZ bosons, where 
decay electrons are reconstructed as
photon candidates, are compared between data and simulation for the 
estimation of photon energy systematic uncertainties.
The residual difference of the mean between data and simulation 
after the energy correction is
considered as the systematic uncertainty due to the energy scale.
The energy resolution uncertainty is estimated by additionally smearing 
photon candidates in simulation
according to the resolution uncertainties of data and simulation.

The systematic uncertainty for unfolding, which comes from the finite size of 
the simulated sample, is
considered when constructing the response matrix. A study based on 
pseudo-experiments is performed for each bin of the response matrix
accounting for the statistical uncertainties of the full simulated sample.
Another variation for the response matrix is performed because of its 
dependence on the shape of the MC spectrum inside the true bins.   
The photon spectra in \pythiahydjet (\PYTHIA) are reweighted 
for the \jetphox photon spectra.
The maximum difference between the nominal and the varied response matrices is 
propagated to the final observables, and
their differences to the nominal values are 
quoted as the systematic uncertainty for unfolding.

Variations of SF obtained from the tag-and-probe method are accounted for 
as a systematic uncertainty of efficiency in the final results.
Photons are measured only with events passing the 
HLT trigger for low-\etg photons with a threshold of 20\GeV for the systematic 
uncertainty of the trigger efficiency. 
The maximum difference between the nominal and the varied efficiencies is 
propagated to the final observables, and
their difference to the nominal values is quoted as the systematic uncertainty for efficiency.

\section{Results}
\subsection{Differential cross section in \texorpdfstring{\pp}{pp} and PbPb collisions}
\label{sec:xSection}
\begin{figure*}[ht]
  \centering
    \includegraphics[width=0.49\textwidth]{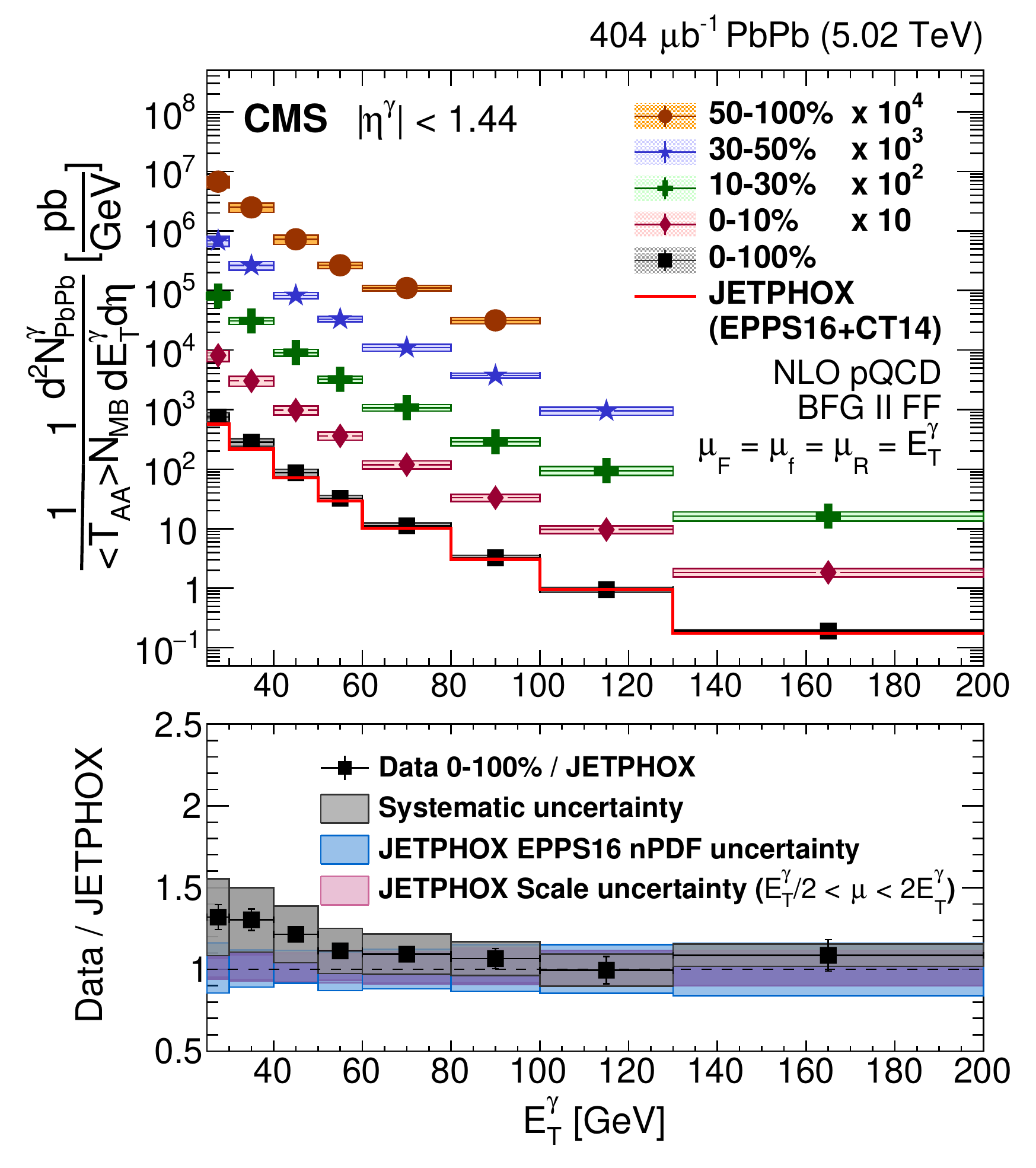}
    \includegraphics[width=0.49\textwidth]{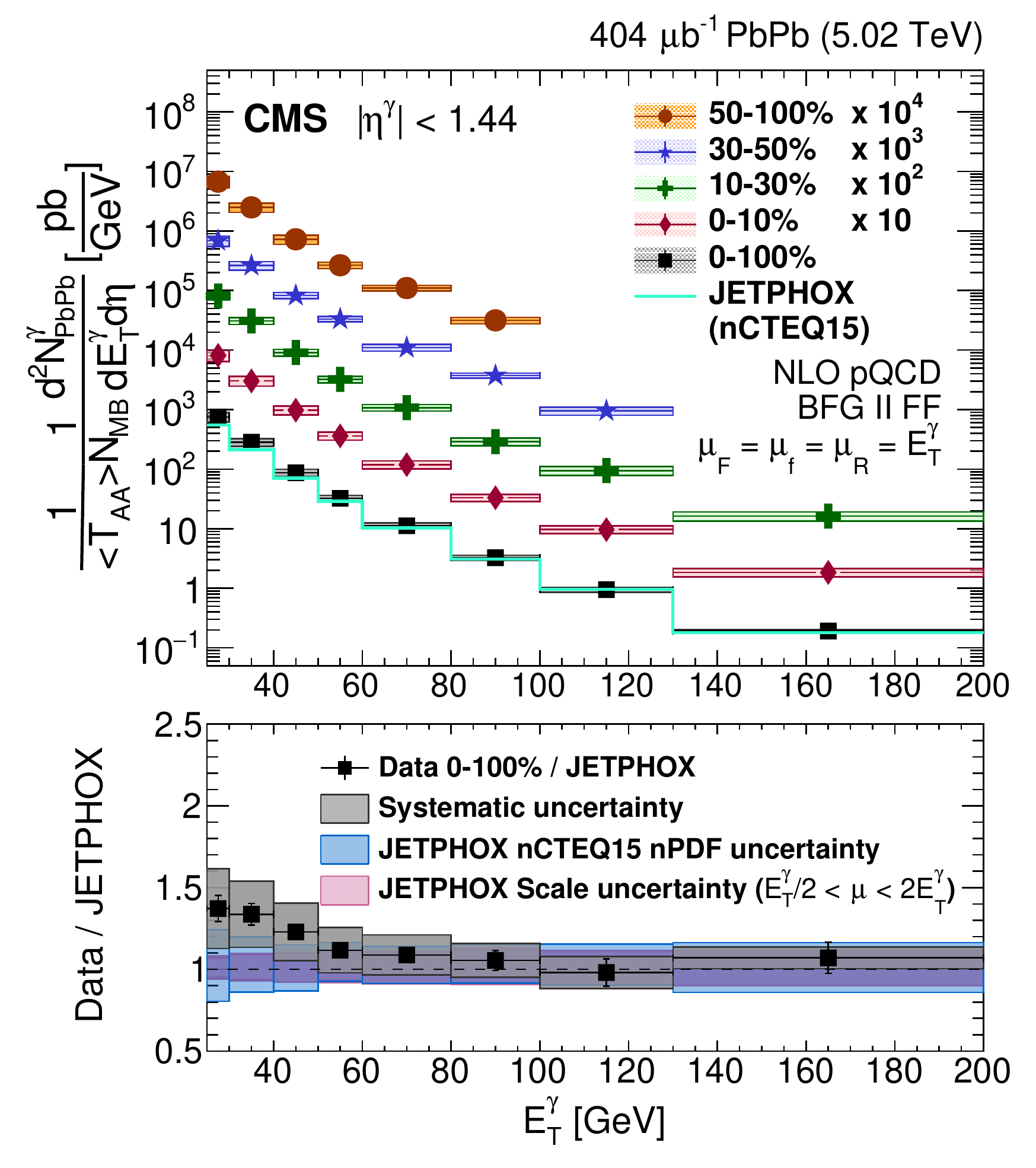}
    \caption{
        Isolated photon spectra (upper) measured as a function of
        \etg for 0--10\%, 10--30\%, 30--50\%, 50--100\%, and 0--100\% PbPb collisions (scaled
        by \TAA) at 5.02\TeV. The spectra are scaled by the factors
        shown in the legend for clarity. 
        The symbols are placed at the center of the bin.
        The vertical bars associated with symbols indicate the statistical
    uncertainties and the horizontal bars reflect the bin width.
        The statistical uncertainties are smaller than the symbols.
        The total systematic uncertainties are shown as boxes in each \etg bin.
        The spectra in the 0--100\% centrality bin are compared to the 
        NLO \jetphox calculations
        with EPPS16+CT14 nPDFs (left) and nCTEQ15 nPDFs (right).
        The ratio of the data in the 0--100\% centrality class to \jetphox is shown in the lower panels. The gray boxes indicate
        the total systematic uncertainties of the data. The blue and red hatched boxes correspond to
        the \jetphox PDF and scale uncertainties, respectively.
    }
    \label{fig:dndet}
\end{figure*}

\begin{figure*}[ht]
  \centering
    \includegraphics[width=0.49\textwidth]{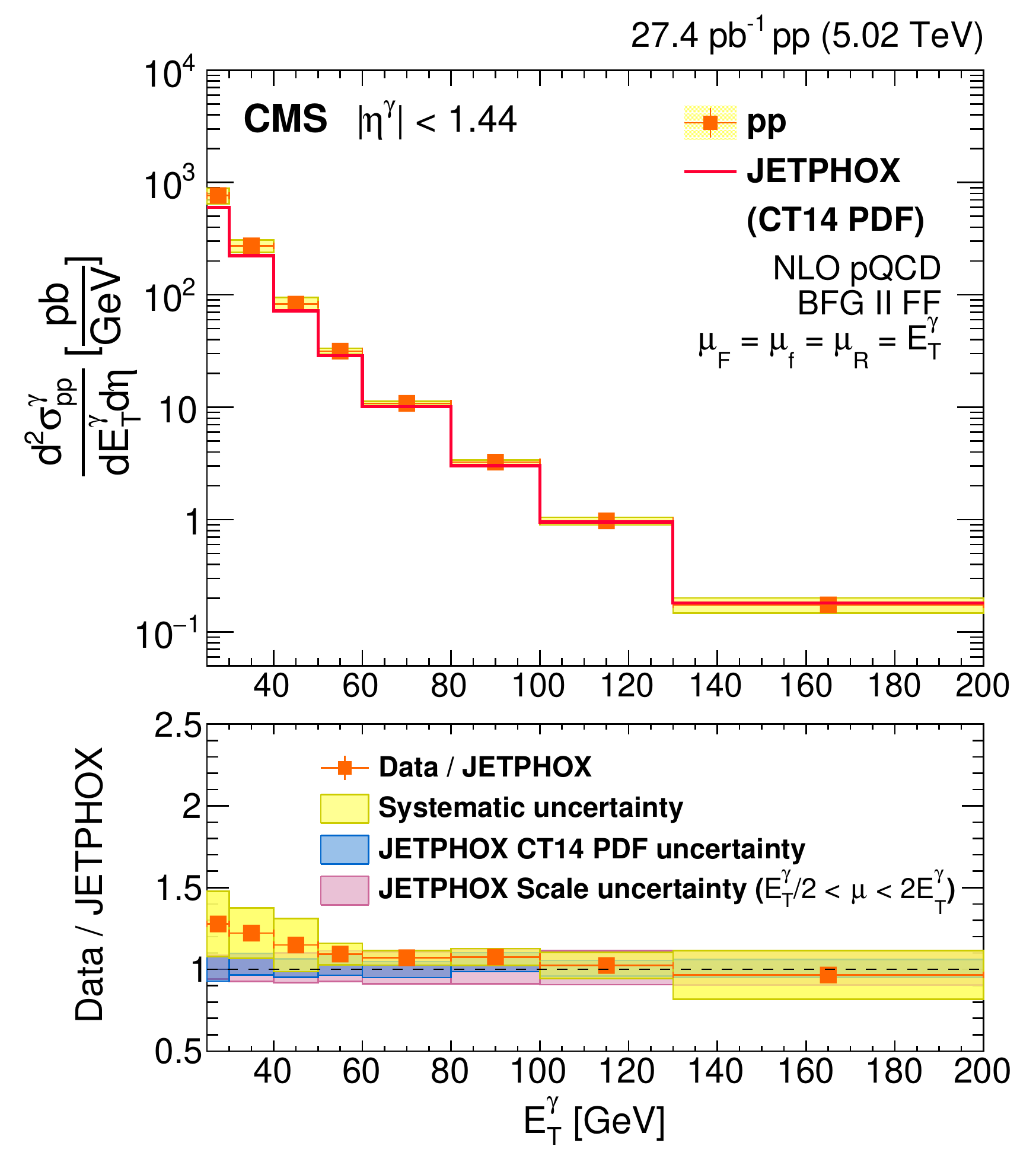}
    \caption{
        Isolated photon cross section (upper) measured as a function of
        \etg in \pp collisions at 5.02\TeV.
        The symbols are placed at the center of the bin.
        The vertical bars associated with symbols indicate the statistical
    uncertainties and the horizontal bars reflect the bin width.
        The statistical uncertainties are smaller than the symbols.
        The total systematic uncertainties are shown as boxes in each \etg bin.
        The data are compared to the NLO \jetphox calculations with CT14 PDFs.
        The ratio of the data to \jetphox is shown in the lower panel. The yellow boxes indicate
        the total systematic uncertainties of the data. The blue and red hatched boxes correspond to
        \jetphox PDF and scale uncertainties, respectively.
    }
    \label{fig:dndet_pp}
\end{figure*}

The \etg-differential cross section scaled by the NN-equivalent 
integrated luminosity per AA collision is defined as
\begin{equation}
\frac{1}{\langle\TAA\rangle}\frac{1}{\nmb}\frac{\rd^2N^{\PGg}_\mathrm{PbPb}}{\rd \etg\rd\eta} = \frac{\ngu}{\langle\TAA\rangle \nmb \Delta \etg \Delta \eta}.
\end{equation}
For the \pp data, the corrected yields are normalized by the integrated luminosity ($\mathcal{L}_{\pp}$) as
\begin{equation}
\frac{\rd^2\sigma^{\PGg}_{\pp}}{\rd\etg\rd\eta}=\frac{\ngu}{\mathcal{L}_{\pp} \Delta \etg \Delta \eta}.
\end{equation}
Figures~\ref{fig:dndet} and~\ref{fig:dndet_pp} show the \etg differential
isolated photon spectra in PbPb collisions for different centrality bins and 
in \pp collisions. The data are compared to the 
NLO pQCD calculations with
\jetphox $v1.3.1\_4$ for MB events.
The CT14~\cite{Dulat:2015mca} PDFs are used for \pp data.
The EPPS16~\cite{Eskola:2016oht} nPDFs based on CT14 PDFs for 
the free-nucleon parton densities (EPPS16+CT14) 
and nCTEQ15~\cite{Kovarik:2015cma} nPDFs are used for PbPb data. 
In the calculations, the BFG set II~\cite{Bourhis:1997yu} is used for the fragmentation function.
The renormalization ($\mu_\mathrm{R}$), factorization ($\mu_\mathrm{F}$) and fragmentation ($\mu_\mathrm{f}$) scales are set to \etg.
Uncertainty in the \jetphox predictions consists of two 
components. First, CT14 PDFs, EPPS16+CT14 nPDFs, and nCTEQ15 nPDFs are varied 
with their 56, 97, and 32 uncertainty sets, respectively. 
The Hessian PDF uncertainties are derived for 90\% confidence level (\CL) 
and scaled down to 68\% \CL~\cite{Buckley:2014ana}. 
Second, the renormalization, factorization, and 
fragmentation scales are varied up and down by a factor 
of two simultaneously. The envelope covered by these variations
is assigned as the scale systematic uncertainty.
As seen in the lower panels of Fig.~\ref{fig:dndet} and~\ref{fig:dndet_pp},
the data are consistent with the \jetphox NLO predictions over the entire  
\etg range 
in both \pp and PbPb collisions, 
considering the quoted statistical and systematic uncertainties.

\subsection{Nuclear modification factors}
\label{sec:raa}
\begin{figure*}[ht]
  \centering
    \includegraphics[width=0.95\textwidth]{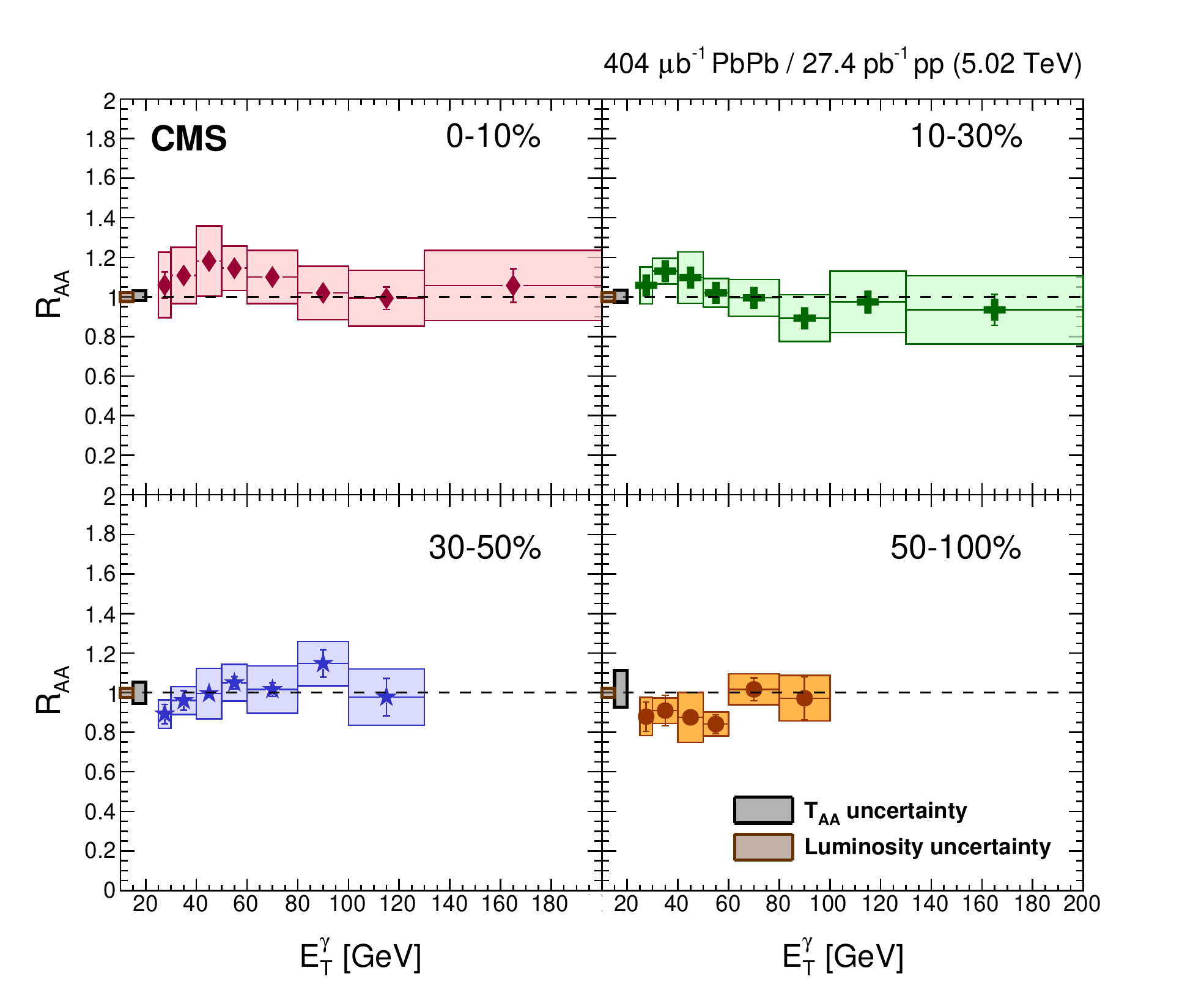}
    \caption{ Nuclear modification factors \raa as a function of
    the photon \etg measured in the 0--10\%, 10--30\%, 30--50\%, and 50--100\% centrality
    ranges in PbPb. 
    The symbols are placed at the center of the bin.
    The vertical bars associated with symbols indicate the statistical
    uncertainties and the horizontal bars reflect the bin width. The total systematic
    uncertainties without the \TAA uncertainty
     are shown as the colored boxes. The \TAA uncertainty,
    common to all points for a given centrality range, is indicated by the gray box
    centered at unity on the left side of each panel. The 2.3\% integrated luminosity 
    uncertainty for \pp data is shown as the brown box at unity at the leftmost position.
    }
    \label{fig:raa_centDep}
\end{figure*}

The nuclear modification factors are calculated by
\begin{equation}
\raa = \frac{1}{\langle\TAA\rangle}\frac{1}{\nmb}\frac{\rd^2 N^{\PGg}_\mathrm{PbPb}/\rd \etg \rd \eta}{\rd^2\sigma^{\PGg}_{\pp}/\rd \etg \rd \eta }.
\end{equation}
Figure~\ref{fig:raa_centDep} shows \raa as a function of 
the isolated photon \etg in different centrality bins. 
The nuclear modification factors exhibit little or no modifications 
of isolated photons in all \etg and centrality bins in PbPb 
collisions, considering the quoted statistical 
and systematic uncertainties.
This indicates that the isolated photons are not
modified by the strongly interacting medium produced in heavy 
ion collisions, which is in contrast to hadrons in PbPb 
collisions~\cite{Khachatryan:2016odn,Sirunyan:2018nsz,Sirunyan:2017isk} 
(\ie $0.3 < \raa < 0.9$ for charged hadrons~\cite{Khachatryan:2016odn} in the same \pt range).

\begin{figure*}[ht]
  \centering
    \includegraphics[width=0.65\textwidth]{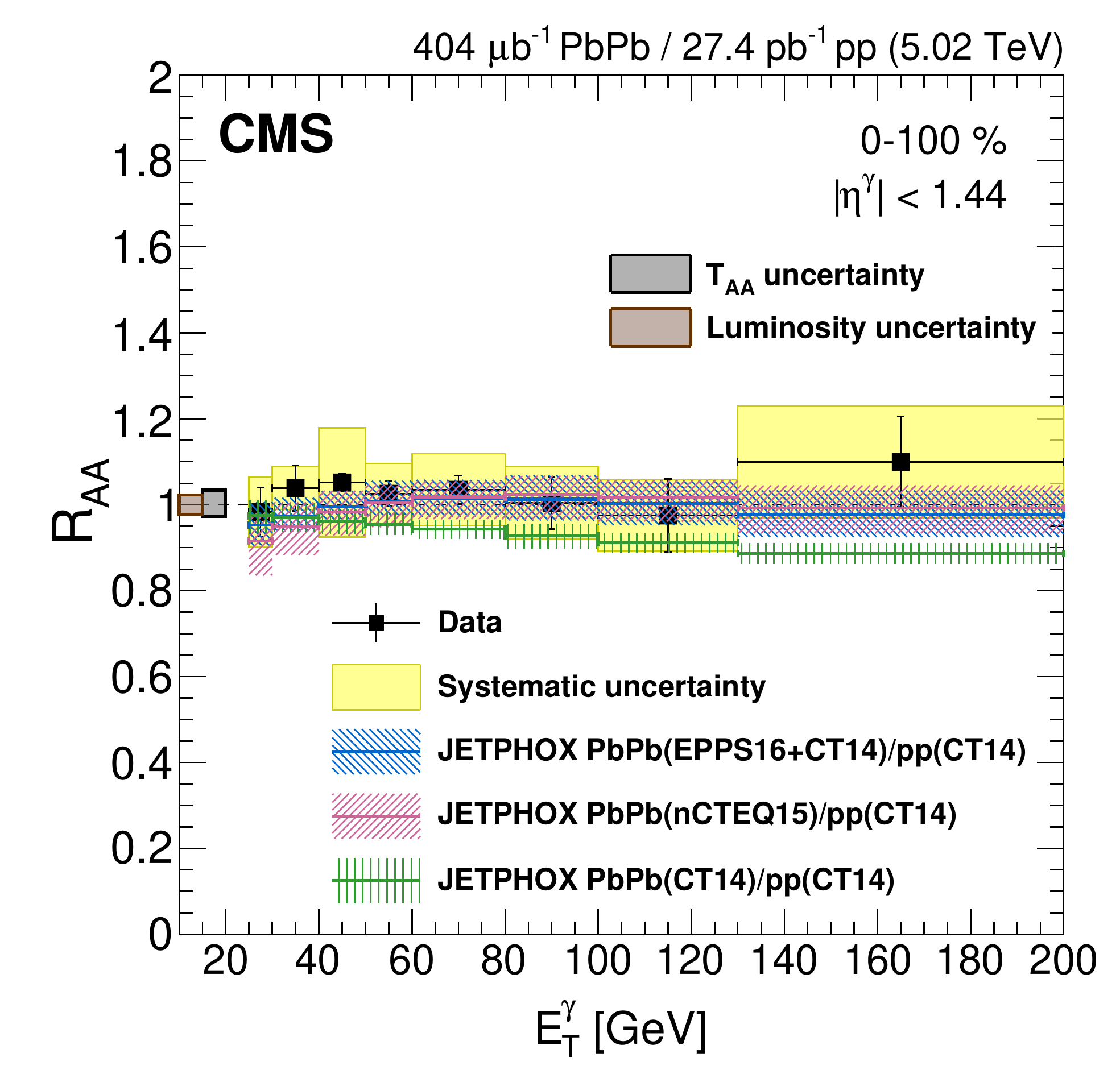}
    \caption{ Nuclear modification factors \raa as a function of
    the photon \etg measured in the 0--100\% centrality
    range in PbPb. 
    The symbols are placed at the center of the bin.
    The vertical bars indicate the statistical
    uncertainties and the horizontal bars reflect the bin width. The total systematic
    uncertainties without the \TAA uncertainty
    are shown by the colored boxes. The 3.4\% \TAA uncertainty,
    common to all points, is indicated by the gray box
    centered at unity on the left side of the panel. The luminosity 
    uncertainty of the \pp data is shown
    as the brown box at unity at the leftmost position.
    The three different NLO \jetphox calculations of 
    EPPS16+CT14 nPDFs, nCTEQ15 nPDFs, and CT14 PDFs for PbPb collisions 
    are divided by the NLO \jetphox calculations with CT14 PDFs for \pp 
    collisions, and compared to the data. 
    The hatched boxes correspond to \jetphox (n)PDF uncertainties.
     }
    \label{fig:raa_0to100}
\end{figure*}

{\tolerance=4800 
The \raa in the inclusive (0--100\%) centrality bin is
compared to the NLO \jetphox calculations with 3 PDFs in Fig.~\ref{fig:raa_0to100}
by taking the ratio of \jetphox predictions for PbPb to that for 
\pp: (EPPS16+CT14)/CT14, nCTEQ15/CT14, and CT14(PbPb)/CT14(\pp).
The CT14(PbPb)/CT14(\pp) ratio shows the isospin effect which is caused 
by the different ratios of u and d quarks in \pp and PbPb collisions.  
The \jetphox scale uncertainties for \raa are canceled in the ratio.
The Hessian PDF uncertainties for \raa are calculated for 68\% \CL.
The \raa measurements are consistent with the \jetphox prediction within the quoted
statistical and systematic uncertainties.
The comparison of data and estimations is limited by the 
uncertainties, barring any firm conclusions for the moment. 
\par}

\section{Summary}

The differential cross sections of photons isolated from nearby particles are reported 
at pseudorapidity $\abs{\eta^{\PGg}}<1.44$ for transverse energy from 25 to 200\GeV 
in proton-proton (\pp) and lead-lead (PbPb) collisions at a center-of-mass energy per 
nucleon pair $\sqrtsNN = 5.02\TeV$ with the CMS detector.
No significant modification of isolated photon cross sections in PbPb 
collisions with respect to scaled \pp collisions 
is observed in the explored kinematic ranges at all collision centralities.
Thus, isolated photons are not affected by the strongly interacting 
medium produced in heavy ion collisions, and they can be a valuable tool 
to access the initial \pt of the associated parton in photon+jet events.

The data are compared with the next-to-leading order perturbative 
quantum chromodynamics calculations using the generator 
\jetphox with CT14 parton distribution functions (PDFs) for \pp data 
and EPPS16 and nCTEQ15 nuclear PDFs for PbPb data.
The predictions are found to be consistent with the 
cross sections for both \pp and PbPb collisions.
The current measurements significantly improve the 
precision compared to the previous CMS results at $\sqrtsNN = 2.76\TeV$ 
and can be valuable inputs for global fits of nuclear PDFs.

\begin{acknowledgments}
  We congratulate our colleagues in the CERN accelerator departments for the excellent performance of the LHC and thank the technical and administrative staffs at CERN and at other CMS institutes for their contributions to the success of the CMS effort. In addition, we gratefully acknowledge the computing centres and personnel of the Worldwide LHC Computing Grid for delivering so effectively the computing infrastructure essential to our analyses. Finally, we acknowledge the enduring support for the construction and operation of the LHC and the CMS detector provided by the following funding agencies: BMBWF and FWF (Austria); FNRS and FWO (Belgium); CNPq, CAPES, FAPERJ, FAPERGS, and FAPESP (Brazil); MES (Bulgaria); CERN; CAS, MoST, and NSFC (China); COLCIENCIAS (Colombia); MSES and CSF (Croatia); RPF (Cyprus); SENESCYT (Ecuador); MoER, ERC IUT, PUT and ERDF (Estonia); Academy of Finland, MEC, and HIP (Finland); CEA and CNRS/IN2P3 (France); BMBF, DFG, and HGF (Germany); GSRT (Greece); NKFIA (Hungary); DAE and DST (India); IPM (Iran); SFI (Ireland); INFN (Italy); MSIP and NRF (Republic of Korea); MES (Latvia); LAS (Lithuania); MOE and UM (Malaysia); BUAP, CINVESTAV, CONACYT, LNS, SEP, and UASLP-FAI (Mexico); MOS (Montenegro); MBIE (New Zealand); PAEC (Pakistan); MSHE and NSC (Poland); FCT (Portugal); JINR (Dubna); MON, RosAtom, RAS, RFBR, and NRC KI (Russia); MESTD (Serbia); SEIDI, CPAN, PCTI, and FEDER (Spain); MOSTR (Sri Lanka); Swiss Funding Agencies (Switzerland); MST (Taipei); ThEPCenter, IPST, STAR, and NSTDA (Thailand); TUBITAK and TAEK (Turkey); NASU (Ukraine); STFC (United Kingdom); DOE and NSF (USA). 
 
  \hyphenation{Rachada-pisek} Individuals have received support from the Marie-Curie programme and the European Research Council and Horizon 2020 Grant, contract Nos.\ 675440, 752730, and 765710 (European Union); the Leventis Foundation; the A.P.\ Sloan Foundation; the Alexander von Humboldt Foundation; the Belgian Federal Science Policy Office; the Fonds pour la Formation \`a la Recherche dans l'Industrie et dans l'Agriculture (FRIA-Belgium); the Agentschap voor Innovatie door Wetenschap en Technologie (IWT-Belgium); the F.R.S.-FNRS and FWO (Belgium) under the ``Excellence of Science -- EOS" -- be.h project n.\ 30820817; the Beijing Municipal Science \& Technology Commission, No. Z191100007219010; the Ministry of Education, Youth and Sports (MEYS) of the Czech Republic; the Deutsche Forschungsgemeinschaft (DFG) under Germany's Excellence Strategy -- EXC 2121 ``Quantum Universe" -- 390833306; the Lend\"ulet (``Momentum") Programme and the J\'anos Bolyai Research Scholarship of the Hungarian Academy of Sciences, the New National Excellence Program \'UNKP, the NKFIA research grants 123842, 123959, 124845, 124850, 125105, 128713, 128786, and 129058 (Hungary); the Council of Science and Industrial Research, India; the HOMING PLUS programme of the Foundation for Polish Science, cofinanced from European Union, Regional Development Fund, the Mobility Plus programme of the Ministry of Science and Higher Education, the National Science Center (Poland), contracts Harmonia 2014/14/M/ST2/00428, Opus 2014/13/B/ST2/02543, 2014/15/B/ST2/03998, and 2015/19/B/ST2/02861, Sonata-bis 2012/07/E/ST2/01406; the National Priorities Research Program by Qatar National Research Fund; the Ministry of Science and Education, grant no. 14.W03.31.0026 (Russia); the Tomsk Polytechnic University Competitiveness Enhancement Program and ``Nauka" Project FSWW-2020-0008 (Russia); the Programa Estatal de Fomento de la Investigaci{\'o}n Cient{\'i}fica y T{\'e}cnica de Excelencia Mar\'{\i}a de Maeztu, grant MDM-2015-0509 and the Programa Severo Ochoa del Principado de Asturias; the Thalis and Aristeia programmes cofinanced by EU-ESF and the Greek NSRF; the Rachadapisek Sompot Fund for Postdoctoral Fellowship, Chulalongkorn University and the Chulalongkorn Academic into Its 2nd Century Project Advancement Project (Thailand); the Kavli Foundation; the Nvidia Corporation; the SuperMicro Corporation; the Welch Foundation, contract C-1845; and the Weston Havens Foundation (USA). \end{acknowledgments}

\bibliography{auto_generated}

\cleardoublepage \appendix\section{The CMS Collaboration \label{app:collab}}\begin{sloppypar}\hyphenpenalty=5000\widowpenalty=500\clubpenalty=5000\vskip\cmsinstskip
\textbf{Yerevan Physics Institute, Yerevan, Armenia}\\*[0pt]
A.M.~Sirunyan$^{\textrm{\dag}}$, A.~Tumasyan
\vskip\cmsinstskip
\textbf{Institut f\"{u}r Hochenergiephysik, Wien, Austria}\\*[0pt]
W.~Adam, F.~Ambrogi, T.~Bergauer, M.~Dragicevic, J.~Er\"{o}, A.~Escalante~Del~Valle, M.~Flechl, R.~Fr\"{u}hwirth\cmsAuthorMark{1}, M.~Jeitler\cmsAuthorMark{1}, N.~Krammer, I.~Kr\"{a}tschmer, D.~Liko, T.~Madlener, I.~Mikulec, N.~Rad, J.~Schieck\cmsAuthorMark{1}, R.~Sch\"{o}fbeck, M.~Spanring, W.~Waltenberger, C.-E.~Wulz\cmsAuthorMark{1}, M.~Zarucki
% ...
\vskip\cmsinstskip
\dag: Deceased\\
1:  Also at Vienna University of Technology, Vienna, Austria\\
% ...
\end{sloppypar}
\end{document}